\documentclass[reprint,aps,prb,showpacs]{revtex4-1}
\usepackage{bbm}
\usepackage{amsfonts}
\usepackage{amsmath}
\usepackage{amssymb}
\usepackage{times}
\usepackage{graphicx}
\usepackage{bm}
\usepackage[colorlinks, linkcolor=blue, anchorcolor=blue, citecolor=red]{hyperref}

\newcommand{\average}[1]{\langle#1\rangle}
\newcommand{\prot}{p_{\text{rot}}}
\begin{document}
\title{Brownian motion in superfluid $^4$He}
\author{Xiao Li$^\dag$}
\email{lixiao@physics.utexas.edu}
\author{Ran Cheng$^\dag$}
\author{Tongcang Li}
\author{Qian Niu$^\S$}
\affiliation{Department of Physics, The University of Texas at Austin, Texas 78712, USA}
\pacs{05.40.Jc, 67.25.de, 42.55.-f}

\begin{abstract}
    We propose to study the Brownian motion of a classical microsphere submerged in superfluid $^4$He using the recent laser technology as a direct investigation of the thermal fluctuations of quasiparticles in the quantum fluid. By calculating the temperature dependence of both the friction coefficient and the strength of the random force, we show that the resonant mode of the fluctuational motion can be fully resolved by the present technology. Contrary to the previous work, it is found that the roton contribution is not negligible, and it even becomes dominant when the temperature is above $0.76$\,K.
\end{abstract}

\maketitle

\begin{figure}[!htb]
   \centering
   \includegraphics[width=0.38\textwidth]{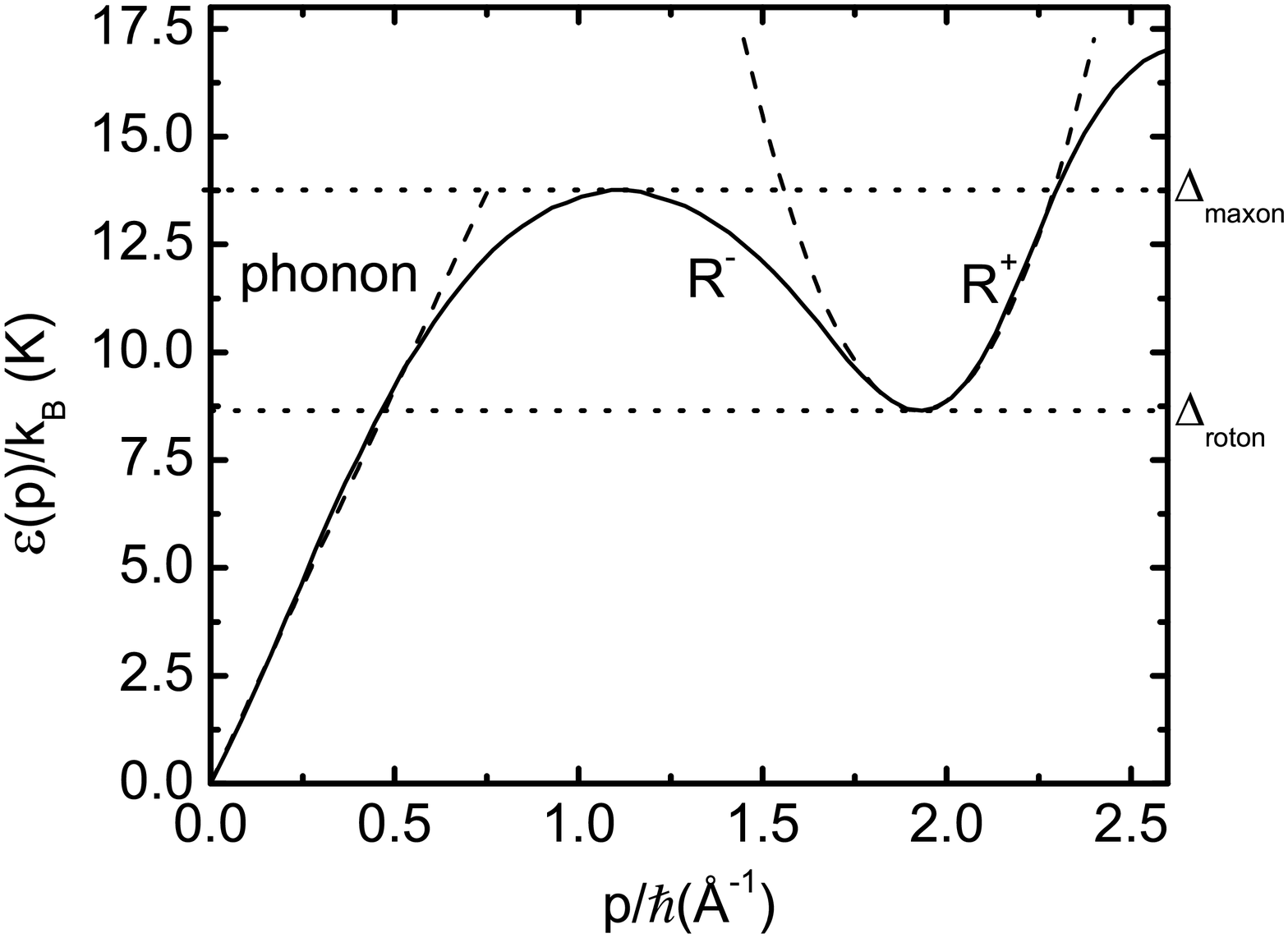}
  \caption{Quasiparticle spectrum in superfluid $^4$He. We have the linear phonon excitations at small momenta and the roton excitations at higher momenta. Rotons with negative and positive slopes are labeled by $R_-$ and $R_+$ respectively. The dashed lines represent the analytical expressions to be used to approximate the spectrum: $\varepsilon(p)=c_s p$ and $\varepsilon(p)=\Delta + \frac{(p-\prot)^2}{2m}$, where $c_s=239$\,m/s, $m=1.06\times 10^{-27}$\,kg, and $\prot/\hbar=1.92$~\AA.\label{Fig:Quasiparticles}}
\end{figure}

Since its discovery in 1827, Brownian motion has attracted broad interest and become a major thrust of statistical physics~\cite{ref:Einstein1,ref:Einstein2,ref:BM1,ref:BM2,ref:BM3}. Driven by thermal fluctuations among the constituent particles, the Brownian motion gave the first direct evidence of the particle nature of the background fluid. However, at extremely low temperatures, quantum effects of the fluid particles become important and certain kinds of quantum fluid emerge, where the system can be described by a macroscopic wave function with long-range phase coherence. In such a case, it is the elementary excitations that play the role of the fluid particles. In thermal equilibrium, those excitations form a gas of quasiparticles that behaves in a similar way to the ordinary atomic gas. They are able to create pressure and resistance when colliding at the interface with a solid. However, their unusual dispersion relations are quite different from those of ordinary particles. A good example is the superfluid $^4$He where two kinds of particle-like excitations, phonons and rotons (see Fig.~\ref{Fig:Quasiparticles}), comprise the viscous part of the fluid~\cite{ref:Landau, ref:Feynman, ref:Quasiparticles}. A natural question arisen is whether these quasiparticles can create Brownian motion. To date, this question has not been addressed by any experiment, and the only theoretical study by Balazs~\cite{ref:Balazs} has ignored the roton contribution without justification. Moreover, Balazs' prediction was far beyond the experimental capability for two reasons: first, the amplitude of the Brownian particle was too small to be monitored at such low temperatures; second, the required observation time was too long due to the tiny restoring force provided by the quartz fiber. But without a fiber, on the experimental side, no object could be suspended in superfluid $^4$He for its small density.

A recent progress in laser technology, however, removes these limitations. It not only promotes the position and time resolutions to a very high level~\cite{ref:RaizenExperiment, ref:FlorinExperiment}, but also is capable of trapping a classical microsphere by laser beams of high intensity that provides a sufficiently strong restoring force, roughly about $10^{6}\sim10^{7}$ times higher than a quartz fiber. In addition, the superfluid medium is transparent to the laser thus the thermal equilibrium of the fluid will not be destroyed by the beams. This technical development therefore makes it possible to test the Brownian motion experimentally in superfluid $^4$He, or potentially, any quantum fluid that does not interact with the laser. Therefore, a complete study of the Brownian motion in superfluid $^4$He, incorporating contributions from both the phonon and the roton, will shed light on both theoretical and experimental investigations.

In this paper, we use the Langevin equation to describe the dynamics of the Brownian particle with the relevant parameters taken from recent experiments~\cite{ref:RaizenExperiment}. The friction coefficient and the strength of the random force are evaluated by the kinetic theory, and expressed as functions of the temperature. We conclude that the Full-Width-at-Half-Maximum~(FWHM) and the mean square of the resonant mode of the fluctuational motion are able to be directly measured by the present experimental techniques~\cite{ref:RaizenExperiment, ref:FlorinExperiment}. Interestingly, while the roton contribution to the above two quantities is negligible at low temperatures, it becomes dominant above $0.76$\,K.

We consider a microsphere with mass $M$ held in superfluid $^4$He by a laser trap with harmonic angular frequency $\omega_0$. Quasiparticles excited by thermal fluctuations in the superfluid will create random forces on the microsphere, resulting in the Brownian motion around its equilibrium position. The motion of the microsphere can be well described by the Langevin equation:
\begin{align}
  M\ddot{\bm{r}}+\gamma\dot{\bm{r}}+M\omega_0^2\bm{r}  = \bm{F}(t),\label{Eq:Langevin}
\end{align}
where $\bm{r}$ denotes the position of the ball, $\gamma$ represents the friction to be estimated below, and $\bm{F}(t)$ is the random force that drives the Brownian motion. It is apparent that different spacial components of $\bm{r}$ are independent in Eq.~\eqref{Eq:Langevin}. Thus in the following we shall only focus on the $z$-component of the Brownian motion. As a consequence, we now consider an effective plate with area $\sigma=\pi r^2$ to represent the microsphere, upon which the problem reduces to one dimension.

We will solve the Langevin equation under the following assumption:
\begin{align}
  \tau_c\ll2\pi/\omega_0\ll\tau,\label{Eq:TimeAssumptions}
\end{align}
which says that the observation time $\tau$ far exceeds the typical period of free oscillation, and the latter is also much larger than the time interval $\tau_c$ between two adjacent collisions from the quasiparticles~\cite{ref:Balazs}. We decompose the random force into a Fourier sum $F_z(t) = \sum_{n}(A_n\cos\omega_nt+B_n\sin\omega_nt)$, where $\omega_n = 2\pi n/\tau$, with $n$ taking integer values, and $A_n = \frac{2}{\tau}\int_{0}^\tau F_z(t)\cos\omega_k t dt$, $B_n = \frac{2}{\tau}\int_{0}^\tau F_z(t)\sin\omega_n t dt$. It is worth mentioning that the Fourier decomposition makes sense only when we regard $F_z(t)$ as a periodic function of time with period $\tau$. By expressing the $z$-component of the displacement as $z(t) = \sum_n z_n(t)$, Eq.~\eqref{Eq:Langevin} can be solved in the frequency domain where
\begin{align}
  z_n^2= \dfrac{1}{2M^2}\dfrac{A_n^2+B_n^2}{(\omega_0^2-\omega_n^2)^2+(\frac{\gamma}{M})^2\omega_n^2}. \label{Eq:FourierComponents}
\end{align}
What subject to direct experimental verifications are the FWHM of the peak and the mean square amplitude of the resonant mode at $\omega_n=\omega_0$, which equal to $\gamma/M$ and $\average{A_0^2+B_0^2}/2\gamma^2\omega_0^2$ respectively, where $\average{\ }$ denotes the ensemble average. The latter one can be brought into a more useful form if we make the following considerations. Because the quasiparticle density is dilute, two adjacent collisions can be considered uncorrelated, and thus we will adopt the white noise assumption that $\average{F_z(t_1)F_z(t_2)}_{\tau} = \mathcal{D}(T)\delta(t_1-t_2)$, where $\mathcal{D}(T)$ is a function of the temperature alone. Also, by regarding $\tau$ as sufficiently large (According to Ref.~\onlinecite{ref:RaizenExperiment}, $\omega_0\sim 2\pi\times10^3$~Hz, thus $\tau=1$~s meets the requirement), we convert the summation $\sum_k$ to the integral $\frac{\tau}{2\pi}\int d\omega$. Then we obtain from Eq.~\eqref{Eq:FourierComponents} the mean square of the resonant mode,
\begin{align}
  \average{z_R^2}= \dfrac{1}{\omega_0^2\pi}\dfrac{\mathcal{D}(T)}{\gamma^2(T)}. \label{Eq:resonantmode}
\end{align}
Therefore, what we need to evaluate are $\gamma(T)$ and $\mathcal{D}(T)$ as functions of the temperature. Although the equipartition relation $\mathcal{D}/4\gamma=k_B T/2$~\cite{ref:BM1, ref:Balazs} obviates the need to calculate both of them, we still do so for strictness.

The friction $\gamma$ originates from the imbalance of the forward and backward scattering when an object has an instantaneous velocity with respect to the background fluid. Suppose $\Delta\sigma$ is a given area on the front side of the effective plate with its normal taken to be the $\hat{z}$-axis. When the plate moves along $z$ direction with velocity $v_z$, the number of quasiparticles within momentum interval [$\bm{p},\,\bm{p}+d\bm{p}$] that is able to collide on $\Delta\sigma$ during time $\Delta t$ ($\tau_c \ll \Delta t \ll 2\pi/\omega_0$) is given by $\Delta\sigma\Delta t d^3\bm{p}|v_z-u_z|N_p/h^3$, where $u_z = \partial\varepsilon(p)/\partial p_z$ is the group velocity of the quasiparticle, and $N_p$ is the number of particles in each unit volume with vector momentum $\bm{p}$. If each collision transfers a momentum $\delta p_z$ to the plate which will be specified below, the resulting force from the front is the average of the total momentum transfer during this time interval divided by $\Delta t$,
\begin{align}
  F_f = \dfrac{\sigma}{h^3}\int d^3\bm{p}\frac{\delta p_z|v_z-u_z|}{e^{\beta\varepsilon}-1}, \quad -\infty<u_z<v_z,
\end{align}
where $\sigma$ is the total area of the plate. Similarly, force $F_b$ from behind is given by the same expression but the range of $u_z$ should be $v_z<u_z<\infty$. Addition of $F_f$ and $F_b$ gives the net resistant force $F_r=F_f+F_b=\gamma v_z+\mathcal{O}(v_z^2)$, with the linear term in $v_z$ being the friction.

\begin{figure}[!htb]
   \begin{center}
   \includegraphics[width=0.39\textwidth]{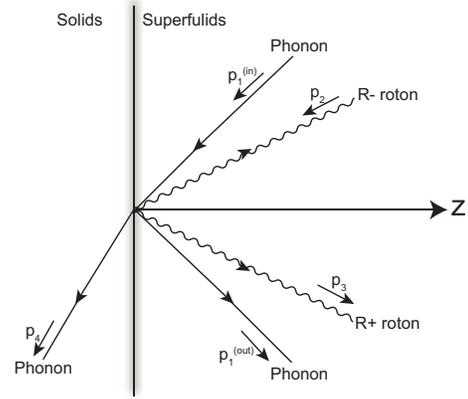}
   \end{center}
  \caption{Reflections and transmissions of quasiparticles on the interface separating the effective plate and the superfluid $^4$He. This diagram only show the case when a phonon with momentum $\bm{p}_1^{\text{(in)}}$ incident on the interface with four outgoing channels: elastic reflection to a phonon $\bm{p}_1$ with possibility $R_{11}$, to a $R_-$ roton $\bm{p}_2$ with possibility $R_{12}$, and to a $R_+$ roton $\bm{p}_3$ with possibility $R_{13}$; inelastic collision that creates a phonon $\bm{p}_4$ into the solid with possibility $R_{14}$. Arrows on each line represents the group velocity of that quasiparticle. Note especially that the group velocity for $R_-$ is in the opposite direction of its momentum. \label{Fig:Reflections}}
\end{figure}

\begin{figure}[!htb]
   \centering
   \includegraphics[width=0.46\textwidth]{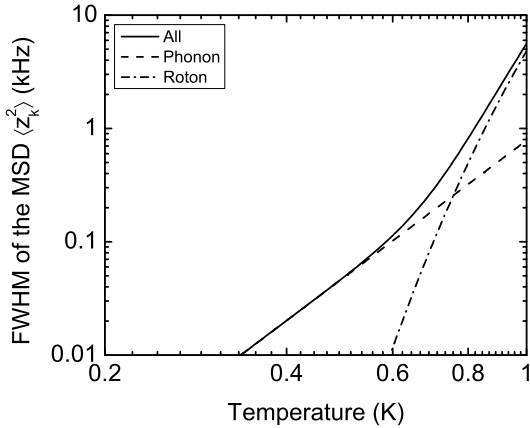}
  \caption{The temperature dependence of the FWHM of the mean square displacement spectra. The thick black line denotes the total width, while the dashed and dash-dotted lines are contributions from the phonon and the roton respectively. The effective area of the plate is taken to be $9\pi/4$ $\mu$m$^2$. \label{Fig:FWHM}}
\end{figure}

Specific computation of $\gamma$ along this line needs detailed information on $\delta p_z$.  Let us now focus on an individual collision that respects the momentum and energy conservations,
\begin{align*}
  p_z+M v_z &= (p_z - \delta p_z)+M v_z', \\
  \varepsilon(p)+\frac12 M v_z^2 &= \varepsilon(\sqrt{p^2-2p_z\delta p_z+(\delta p_z)^2})+\frac12 M v_z'^2,
\end{align*}
and we regard the mass of the Brownian particle $M$ as sufficiently large ($M=2.8\times10^{-14}$~kg experimentally) so that each process is a hard wall collision. Problem arises when the energy of the incoming quasiparticle $\varepsilon(p)$ lies in the region $(\Delta_{_\text{roton}}, \Delta_{_\text{maxon}})$, for which the above equations admit three inequivalent solutions of $\delta p_z$. That is to say, for example, for a phonon carrying definite momentum and energy incident on the plate, it has three different outgoing channels labeled by $i=1,2,3$ representing the phonon, $R_-$ and $R_+$ rotons respectively. We may also add the possibility of inelastic collision that creates phonons in the Brownian particle labeled as $i=4$ ~\cite{ref:Reflections1, ref:Reflections2}. This multi-channel collision process is illustrated in Fig.~\ref{Fig:Reflections}. The momentum transfer $\delta p_z$ then depends on the transition probability $R_{ij}$ connecting the $i$-th and the $j$-th channels. However, it is remarkable that the final expression of $\gamma$ turns out to be independent of $R_{ij}$ as if there were no inter-channel transitions. We omit the detailed argument of this result, as it shares similar logic with the problem of quasiparticle pressure in superfluid $^4$He~\cite{ref:Reflections1}, where all inter-channel transitions mutually cancel. After some manipulations, both the phonon and the roton contributions can be brought into the same form as
\begin{align}
  \gamma(T)=\frac{4\pi\sigma}{h^3} \int_0^{\infty} \mathrm{d}p \frac{p^3}{e^{\beta \varepsilon(p)}-1}.
\end{align}
Insertion of the dispersion relations $\varepsilon(p)=c_s p$ and $\varepsilon(p)=\Delta + \frac{(p-\prot)^2}{2m}$ (See Fig.~\ref{Fig:Quasiparticles}) yields the contributions from the phonon and the roton respectively,
\begin{align}
  &\gamma_{\text{ph}}(T) = \dfrac{\sigma\pi^2}{30\hbar^3c^4}(k_{_B}T)^4,\\
  &\gamma_{\text{rot}}(T) =\dfrac{\sigma \prot^3}{\hbar^3\pi^{\frac32}}\sqrt{\frac m2} e^{-\frac{\Delta}{k_{_{B}}\!T}}\sqrt{k_{_{B}}\!T}\left(1+\dfrac{3mk_{_{B}}\!T}{\prot^2}\right), \label{Eq:Viscosity}
\end{align}
and the total friction coefficient is $\gamma(T) = \gamma_{\text{ph}}(T)+\gamma_{\text{rot}}(T)$. For a Brownian particle with $M=2.8\times10^{-14}$~kg~\cite{ref:RaizenExperiment}, the FWHM $\gamma(T)/M$ is depicted in Fig.~\ref{Fig:FWHM}. The crossing point of $\gamma_{\text{ph}}(T)$ and $\gamma_{\text{rot}}(T)$ is at 0.76\,K. While the roton contribution is negligible at low temperatures, it becomes dominate above 0.76\,K. We see from the figure that the typical FWHM ranges from 10\,Hz to 1\,kHz, and that the departure from pure phonon contribution above 0.76\,K is of the order of 1\,kHz. These are well within the capability of current experiments where the resolution of frequency is down to 1\,Hz.

Having fully evaluated the friction coefficient $\gamma(T)$, now we turn to the more involved quantity $\mathcal{D}(T)$ which comes from the fluctuation of the random force exerted on the Brownian particle. Mathematically, the fluctuation is embodied in the statistical deviation of the quasiparticles distribution $n(\bm{r},\bm{p})$ in the six-dimensional phase space which satisfies $N_p=\int d^3\bm{r}n(\bm{r},\bm{p})$. Neglecting the influence on $n(\bm{r},\bm{p})$ of the scattering among quasiparticles, we again assume that $n(\bm{r},\bm{p})$ on different phase points are independent so that $\average{n(\bm{r},\bm{p})n(\bm{r'},\bm{p'})}\sim\delta^3(\bm{r}-\bm{r'})\delta^3(\bm{p}-\bm{p'})$, i.e., only contributions from the same phase point are kept. Then with the help of the Bose relation $\average{(N_p-\bar{N_p})^2}=\bar{N_p}+\bar{N_p}^2$, we know after some manipulations the total fluctuation of the momentum transfer during $\Delta t$ is: $\average{G_z^2}_{\!_{\Delta t}} = \sigma\Delta t \int d^3\bm{p}/h^3|u_z|(\delta p_z)^2(\bar{N_p}+\bar{N_p}^2)$. The corresponding fluctuation of the random force should be $\average{F_z^2}_{\!_{\Delta t}}=\average{G_z^2}_{\!_{\Delta t}}/(\Delta t)^2$ and the quantity of central interest is $\mathcal{D}=\average{F_z^2}_{\!_{\Delta t}}\Delta t$, which gets rid of the $\Delta t$ dependence,
\begin{align}
  \mathcal{D}(T) = \dfrac{\sigma}{h^3}\!\int\! d^3\bm{p}\left|\frac{\partial\varepsilon(p)}{\partial p_z}\right|\dfrac{(\delta p_z)^2e^{\beta\varepsilon(p)}}{(e^{\beta\varepsilon(p)}-1)^2}. \label{Eq:DT}
\end{align}
Again, by inserting the dispersion relations, a straightforward calculation leads us to:
\begin{align}
  &\mathcal{D}_{\text{ph}}(T) = \dfrac{\sigma\pi^2}{15\hbar^3 c^4}(k_{_B}T)^5,\\
  &\mathcal{D}_{\text{rot}}(T) =\dfrac{\sigma \prot^3 \sqrt{2m}}{\hbar^3\pi^{\frac32}}e^{-\frac{\Delta}{k_{_{B}}\!T}}(k_{_{B}}\!T)^{\frac32}\left(1+\dfrac{3mk_{_{B}}\!T}{\prot^2}\right), \label{Eq:TotalAmplitude}
\end{align}
and $\mathcal{D}(T) = \mathcal{D}_{\text{ph}}(T)+\mathcal{D}_{\text{rot}}(T)$. We mention in passing that the equipartition relation $\mathcal{D}/4\gamma=k_B T/2$ does hold separately for phonon and roton excitations.

\begin{figure}[!htb]
   \centering
   \includegraphics[width=0.46\textwidth]{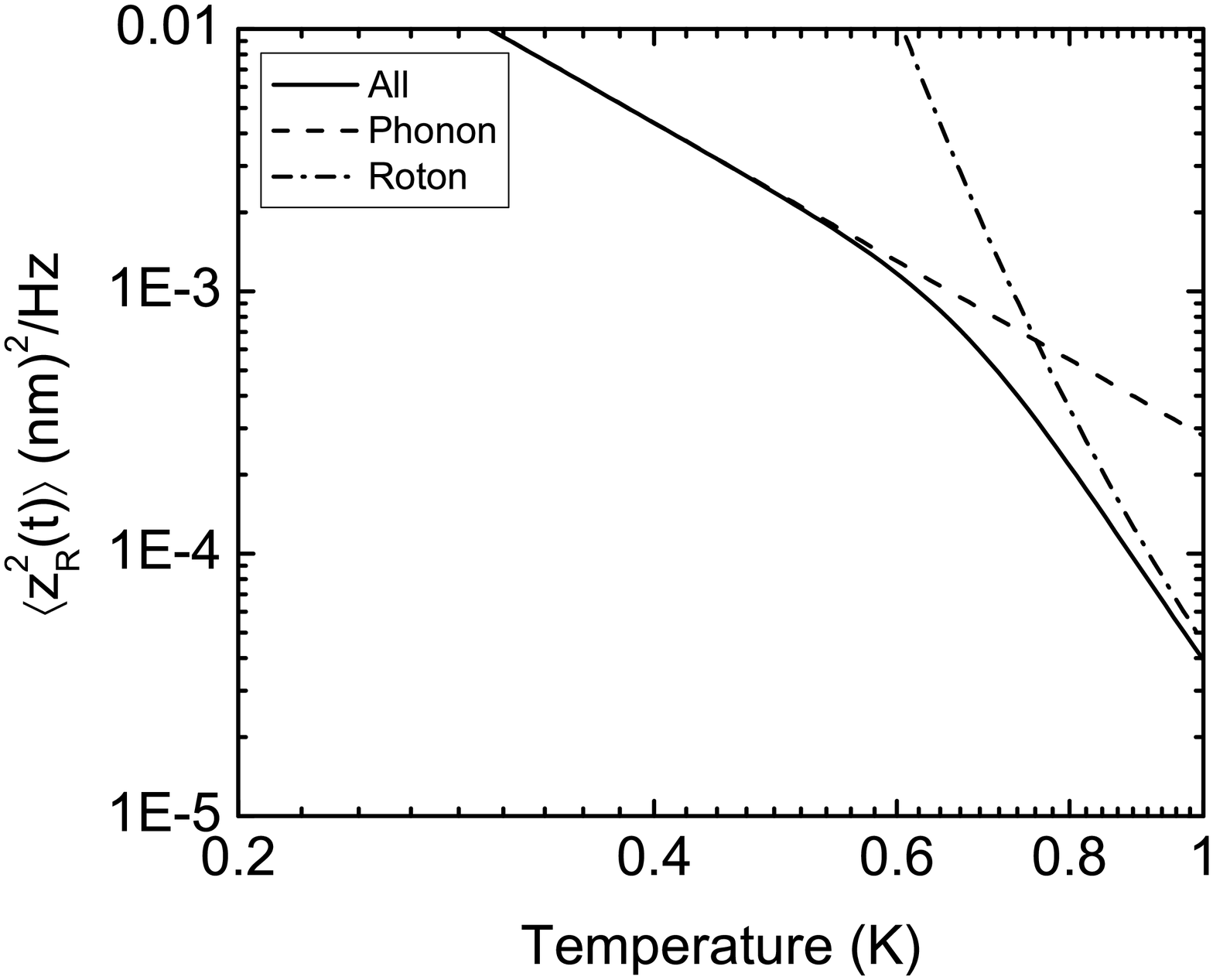}
  \caption{The temperature dependence of the square amplitude of the resonant mode. The thick line represents the result of Eq.~\eqref{Eq:Amplitude}. In comparison, we also plotted the result in existing literatures by the dashed line, where the roton contribution is absent. The dash-dot line depicts the case if the roton contributes alone. Parameters relating to experiment are the same as those in Fig.~\ref{Fig:Reflections}, and the resonant frequency $\omega_0$ is taken to be $2\pi\times6000$\,Hz. \label{Fig:Resonant}}
\end{figure}

Equipped with the friction coefficient and the fluctuation of random force, we are able to evaluate the temperature dependence of the resonant mode, which serves as another quantity for direct experimental test. In view of the equipartition relation, the resonant mode in Eq.~\eqref{Eq:resonantmode} is
\begin{align}
  \average{z_R^2}= \dfrac{2 k_B T/\omega_0^2\pi}{\gamma_{\text{ph}}(T)+\gamma_{\text{rot}}(T)}. \label{Eq:Amplitude}
\end{align}
This is plotted in Fig.~\ref{Fig:Resonant}, where $0.76$\,K is again identified as the turning temperature. The higher the temperature, the lower the amplitude of the resonant mode is. Fortunately, the lowest $\sqrt{\average{z_R^2}}$ seen from the plot is about $10^{-3}\sim10^{-2}$\,nm/$\sqrt{\text{Hz}}$, far beyond the experimental resolution $3.9\times10^{-5}$\,nm/$\sqrt{\text{Hz}}$~\cite{ref:RaizenExperiment}. At low temperatures, however, the resonant mode is dominated by the phonon excitation and diverges as $T^{-3}$. This seemingly counterintuitive result is resolved when we consider the assumption made in Eq.~\eqref{Eq:TimeAssumptions}, which has been discussed in Ref.~\onlinecite{ref:Balazs}. As the temperature goes down, $\tau_c$ sharply increases and the above assumption becomes invalid, where a new theory is required. Fortunately, for a typical angular frequency $\omega_0\sim 2\pi\times6000$~Hz, Eq.~\eqref{Eq:TimeAssumptions} holds until the temperature is lowered to $10^{-6}$~K.

Before conclusion, two further comments are in order. First, the motion of the Brownian particle will transfer kinetic energy to the background fluid. As a consequence the mass $M$ in Eq.~\eqref{Eq:Langevin} should be the effective mass instead of the bare mass $m_0$. Below $T=$1\,K, it can be approximated as~\cite{ref:EffectiveMass}:
$M=m_0+\dfrac 12 \rho_s\times\dfrac43\pi r^3,$
where $\rho_s$ is the superfluid density and $r$ is the radius of the microsphere. For the microsphere with a diameter of 3\,$\mu$m and mass of $m_0=2.8\times10^{-14}$\,kg, the correction term is roughly about 5\%.

Moreover, the white noise assumption on the random force as well as the delta correlated distribution $n(\bm{r},\bm{p})$ implies the independence of different quasiparticle modes. This is quite reasonable when the density of quasiparticles is dilute, as the relative strength of quasiparticle scattering is proportional to its square. At roughly $T=$1~K, the percentage of the normal fluid formed by the quasiparticles is less than 5\%~\cite{ref:Peshkov}, and we would expect a negligible effects of quasiparticle scattering.

In conclusion, we have studied the Brownian motion of a classical microsphere driven by thermally excited quasiparticles in superfluid $^4$He. Contrary to previous work, we claim the importance of contributions from both the phonon and the roton excitations, and find the turning temperature of their relative importance at 0.76\,K. More importantly, the two predictions we give on the FWHM and the resonant mode are able to be tested in current experiments. Generalization to other types of quantum fluids are left for future inquires.\\

The authors are grateful to Prof.~M.~Raizen for helpful discussions. This work is supported by NSF (Grant No.~DMR0906025), DOE (Grant No.~DE-FG03-02ER45958, Division of Materials Science and Engineering), and the Welch foundation (Grant No.~F-1255).\\

$^\dag$ These authors contribute equally to this work.

$^\S$ On leave from the University of Texas at Austin.\\

\end{document}